\documentclass[pra,aps, twocolumn,
showpacs]{revtex4}
\usepackage{epsfig}
\usepackage{graphicx}

\begin{document}
\title{Microwave whispering gallery resonator for efficient optical up-conversion}

\author{D.\ V. Strekalov$^1$}
\author{H.\ G.\ L. Schwefel$^2$}
\author{A.\ A. Savchenkov$^3$}
\author{A.\ B. Matsko$^3$}
\author{L.\ J. Wang$^2$}
\author{N. Yu$^1$}
\affiliation{$^1$Jet Propulsion Laboratory, California Institute of
Technology, 4800 Oak Grove Drive, Pasadena, California 91109-8099\\
$^2$Max-Planck-Institute for the Science of Light, D-91058 Erlangen, Germany\\
$^3$OEwaves Inc., 2555 East Colorado Boulevard, Pasadena, CA 91107}

\date{\today}

\begin{abstract}
Conversion of microwave radiation into the optical range has been predicted to reach unity quantum efficiency in whispering gallery resonators made from an optically nonlinear crystal and supporting microwave and optical modes simultaneously. In this work we theoretically explore and experimentally demonstrate a resonator geometry that can provide the required phase matching for such a conversion at any desired frequency in the sub-THz range. We show that such a ring-shaped resonator not only allows for the phase matching, but also maximizes the overlap of the interacting fields. As a result, unity-efficient conversion is expected in a resonator with feasible parameters.
\end{abstract}


\maketitle

\section{Introduction}
Nonlinear frequency conversion of far-infrared or microwave signals into the optical domain has been actively used for detection of such signals \cite{chiou72apl,abbas76ao,albota04ol,karstad05ole,temporao06ol,vandevender07josab,ding06,khan07,strekalov08THz1}.  The relative ease of optical signal detection compared to e.g. those in the sub-THz range, in combination with an intrinsically noiseless character of nonlinear frequency conversion, explains the close attention this method has been receiving. Its main drawback, however, is its low conversion efficiency. The highest power conversion efficiency known to us is about 0.5\%, which has been only recently achieved for 100 GHz signal using 16 mW of CW optical pump at 1560 nm \cite{strekalov08THz1}. This number corresponds to the photon-number conversion efficiency of approximately $2.6\cdot 10^{-6}$, as follows from the Manley-Rowe relation.

Highly efficient and noiseless upconversion of microwave radiation into the optical domain would open up numerous possibilities in microwave imaging and communications. One example of such a possibility, that we have discussed earlier \cite{strekalov08THz1}, is microwave photon counting at room temperature. Reaching the photon-counting regime in the sub-THz or THz range would be an achievement important for quantum information processing and computing, sub-millimeter spectroscopy and astronomy, security, and for other areas where the ultimate sensitivity detection of microwave radiation is desired. Unfortunately, even the most efficient up-converter to-date is still seven orders of magnitude short of the photon-counting regime for 100 GHz photons at room temperature \cite{strekalov08THz1}. On the other hand, the theoretical analysis \cite{Matsko08THzTheory} shows that this regime can be achieved. The key to its realization is reaching unity conversion efficiency. An all-resonant whispering gallery mode (WGM) configuration of the frequency converter should achieve this goal. 

WGM resonators with optical nonlinearity have been successfully used in nonlinear optics in general and in microwave photonics in particular
\cite{cohen01el-a,rabiei02jlt,ilchenko03mod,savchenkov09ssb,hosseinzadeh06mtt}. In a recent experiment \cite{strekalov08THz1} a lithium niobate WGM resonator with the optical free spectral range (FSR) $\Omega=2\pi\cdot 12.64$ GHz was irradiated by a microwave signal with the frequency near $\omega_M=8\cdot\Omega\approx 2\pi\cdot 101.12$ GHz. This device is analogous to a lower-frequency electro-optical modulator \cite{cohen01el-a,savchenkov09ssb,ilchenko03mod},
except that the microwave signal excites the sidebands not in the
pair of adjacent to the carrier optical WGMs, but in those across
eight of optical FSRs. 

\section{General theoretical consideration}
To describe the operation of such a modulator, we follow the steps of \cite{ilchenko03mod} and introduce the interaction Hamiltonian
\begin{equation}
\hat{H_i}=\hbar g(\hat{b}^\dagger_-\hat{c}^\dagger\hat{a}+\hat{b}_+\hat{c}^\dagger\hat{a}^\dagger)+c.c.,\label{H}
\end{equation}
which couples the optical pump and microwave signal WGMs (photon annihilation operators $\hat{a}$ and $\hat{c}$, respectively) with the Stokes and anti-Stokes optical upconverted signal ($\hat{b}_-$ and $\hat{b}_+$, respectively). In case of a single side-band modulation \cite{savchenkov09ssb} only one, either the Stokes or the anti-Stokes, term is present in Hamiltonian (\ref{H}).

The coupling constant
\begin{equation}
g=\omega_0r_{ij}\frac{n_an_b}{n_c}\sqrt{\frac{\pi\hbar\omega_c}{2V_c}}
\left(\frac{1}{V}\int_V\,dV\Psi_a^*\Psi_b\Psi_c\right)
\label{g}
\end{equation}
includes the (bracketed) factor describing the overlap between the fields inside the resonator. In Eq.~(\ref{g}), $\omega_a\approx\omega_b\equiv\omega_0$ is the optical frequency, $n_{a,b}$ are the refraction indices and $V_a\approx V_b\equiv V$ is the optical mode volume; $\omega_c$, $n_c$ and $V_c$ are similar microwave parameters. The effective electro-optical coefficient $r_{ij}$ is determined by the fields configuration.

Irradiating the resonator with the microwaves by placing the former near a waveguide opening is a very inefficient method of coupling the microwaves with the optical WGMs, as the main part of the microwave energy reflects back into the waveguide or scatters into space. This problem can be solved by using an all-resonant frequency up-converter, supporting microwave WGMs as well as optical ones. Experimental studies of microwave WGMs in crystalline disks and rings carried out in the 1980's shown remarkable quality factors of $Q\approx 10^{10}$ \cite{braginsky87}.

For the all-resonant up-converter, all wave functions $\Psi$ in (\ref{g}) are optical or microwave eigenfunctions of the WGM resonator. The eigenfrequencies $\omega_a(L_a)$ of the main family of the optical WGMs is found \cite{ilchenko03mod} as
\begin{equation}
\omega_a=\frac{cL_a}{Rn_a},\label{eigen}
\end{equation}
where $L_a$ is the orbital momentum for this WGM, and $R$ is the resonator radius. Similar equations hold for the Stokes and anti-Stokes frequencies $\omega_{b_\pm}$. 

The Hamiltonian (\ref{H}) and Eq. (\ref{g}) lead to the following phase-matching conditions, that are essentially the energy and angular momentum conservation equations for the anti-Stokes and Stokes processes:
\begin{equation}
\omega_{b_\pm}=\omega_a\pm\omega_c
\quad{\rm and}\quad
L_{b_\pm}=L_a\pm L_c,\label{phasematch}
\end{equation}

Subtracting Eq.~(\ref{eigen}) for $\omega_a$ from that for $\omega_{b_\pm}$ and making substitution (\ref{phasematch}), we find the phase-matching equation
\begin{equation}
\omega_c(L_c)=\Omega_b L_c\pm \omega_a(n_a-n_b)/n_b\label{phasematch1}
\end{equation}
for the anti-Stokes (plus) and Stokes (minus) conversion processes. In (\ref{phasematch1}) $\Omega_b= c/(Rn_b)$ is the optical FSR for the signal WGM. We have neglected its frequency dispersion replacing $n_{b_\pm}\equiv n_b$, but kept the distinction between $n_a$ and $n_b$ which is due to the birefringence. 

The phase matching condition (\ref{phasematch1}) needs to be solved jointly with the microwave dispersion equation, which depends on the resonator geometry. 
Previously \cite{strekalov09lasphys} we have observed lower frequency microwave WGMs in lithium niobate disks in the 30 GHz range.  
One important conclusion of that study has been that the microwave WGMs in a disk resonator have poor spatial overlap with the optical modes that occupy just a few tens of microns near the disk rim. This greatly reduces the overlap factor in (\ref{g}). Therefore we decided to use ring resonators, made so that the optical axis of the crystal is parallel to the ring axis. When filled with a low-constant dielectric, such a ring tends to concentrate the microwave field inside, enforcing a better overlap with the optical WGMs.

Once the phase-matching conditions are fulfilled, the microwaves-to-optics conversion will be efficient as long as the nonlinear coupling rate exceeds the loss rate in the microwave WGM, that is, if the total loss rate of the THz
mode $\gamma=\gamma_{nl}+\gamma_{abs}$ is dominated by the rate of
nonlinear frequency conversion $\gamma_{nl}$. Experimentally this means that the microwave WGM resonances will be considerably broadened when the optical pump is turned on. Then if 
the optical WGMs couple strongly to the input and output free-space beams (i.e. the resonator is optically over-coupled) the unity-efficient conversion is theoretically possible \cite{Matsko08THzTheory}.

Efficient in- and out-coupling of the optical WGMs requires the external rim of the ring to be shaped as a spheroid with the radii ratio equal to \cite{strekalov08THz1}
\begin{equation}
\frac{\rho}{R} = \frac{n_p^2-n^2}{n_p^2},\label{r2r}
\end{equation}
where $n_p$ is the refraction index of the prism used for the optical coupling. Let us assume for the rest of the paper that the optical pump has the wavelength $\lambda= 1.55$ $\mu$m ($\omega_0=1.2\cdot 10^{15}$ s$^{-1}$) . Then for a diamond prism $n_p=2.384$, while e.g. for lithium niobate $n_{a,b} = n_e = 2.138$. Therefore according to (\ref{r2r}), $\rho/R=0.196$. 

\section{Type-I up-conversion}
\begin{figure}[b]
\vspace*{-0.2in}
\centerline{
\input epsf
\setlength{\epsfxsize}{3.3in} \epsffile{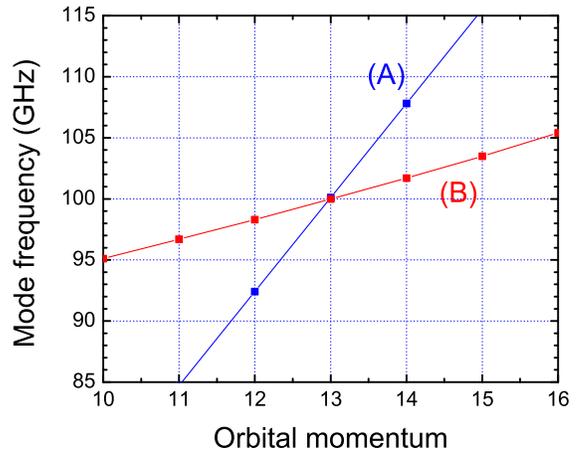}}\vspace*{-0.2in}
\caption[]{\label{fig:phasematch} A phase-matching solution for lithium niobate, Type-I up-conversion, is achieved at $\omega_c= 2\pi\cdot 100$ GHz, $L_c=13$. Curve (A) is the phase-matching curve resulting from Eq.~(\ref{phasematch1}); curve (B) is the microwave dispersion curve calculated numerically. }\vspace*{-0.1in}
\end{figure}

Let us find the phase matching solutions. First, we assume that the microwave as well as \emph{both} optical fields are polarized along the optical axis, so that the largest nonlinearity coefficient of lithium niobate $d_{33}$ is used. In this configuration, which we dub Type-I up-conversion, $n_a=n_b$ and the last term in Eq. (\ref{phasematch1}) disappears. To find the phase-matching solutions, we plot Eq.~(\ref{phasematch1}) together with the microwave dispersion curve, see Fig.~\ref{fig:phasematch}.
Intersection of the two curves at an integer value of $L_c$ indicates that the phase-matching is achieved. The accuracy to which the two frequencies should match at this value of $L_c$ is determined by the smaller of the optical and microwave WGM linewidths. Notice that the microwave dispersion curve (B) in Fig.~\ref{fig:phasematch} depends on all four geometric parameters of the ring, while the phase-matching curve (A) depends only on the external radius $R$. This gives us enough freedom to achieve the phase-matching at a desired microwave frequency. In the example shown in  Fig.~\ref{fig:phasematch} the phase-matching was found for $\omega_c= 2\pi\cdot 100.0$ GHz at $L_c=13$. The ring thickness was $h=292\,\mu$m, its inner and outer radii were $R_{in}=2.48$ mm and $R=2.9$ mm, and the rim curvature found from (\ref{r2r}) was $\rho=568\,\mu$m.

\begin{figure}[t]
\begin{center}
\includegraphics[clip,angle=0,height=5cm]{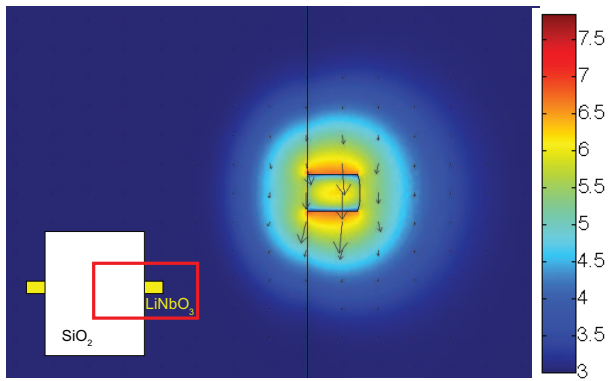}
\caption{Absolute value of the $E_z$-component of a microwave WGM mode in a lithium niobate resonator with $\omega_c= 2\pi\cdot 100$ GHz, $L_c=13$. The plot is on a log scale with color map given on the right.  A sketch of the geometry is shown in the inset, where the red box represents the calculation window.}
\label{fig:wfRing}
\end{center}\vspace*{-0.2in}
\end{figure}

The microwave dispersion curve is obtained by numerical simulation. These simulations have been carried out in a finite element solver, COMSOL \cite{comsol} adapting an axis symmetrical formulation by Oxborrow \cite{oxborrow}.
Fused silica was selected as the post material because of its low microwave absorption and relatively small microwave refraction index (n=1.9 at 100 GHz), compared to the larger values $n_e=5.15$ and $n_o=6.72$ for lithium niobate \cite{palik}.
In this configuration the microwave field is strongly concentrated inside of the resonator ring, see Fig.~\ref{fig:wfRing}, which provides both 
good coupling with the external waveguide and good overlap with the optical field. 

We would like to point out that the purpose of the numeric simulation is not to achieve the exact phase-matching, but only to unambiguously determine the angular momentum $L_c$. An error small compared to the microwave FSR would be tolerable, because the phase matching can be fine-tuned via modifying the microwave dispersion of the system. For example, placing a metal ring on the fused silica post near the resonator ring, we have been able to tune the microwave frequency by more than 0.5 GHz without appreciable degradation of the quality factor. 

\section{Type-II up-conversion}
Now let us consider the Type-II up-conversion, i.e. when the signal polarization is orthogonal to the pump polarization \cite{savchenkov09ssb}. In this case the birefringence term in (\ref{phasematch1}) does not vanish. In fact, in strongly birefringent materials it can be quite large. In lithium niobate, for example, it is equal to approximately $\pm 2\pi\cdot 6.6$ THz. The ``plus" sign corresponds to either anti-Stokes conversion of the ordinary polarized pump, or Stokes conversion of the extraordinary polarized pump. Similarly, the ``minus" sign corresponds to either Stokes conversion of the extraordinary polarized pump, or anti-Stokes conversion of the ordinary polarized pump. It is easy to see that neither situation allows for the phase matching solution compatible with the microwave dispersion equation and lying in the transparency window of lithium niobate.

\begin{figure}[t]
\vspace*{-0.2in}
\centerline{
\input epsf
\setlength{\epsfxsize}{3.3in} \epsffile{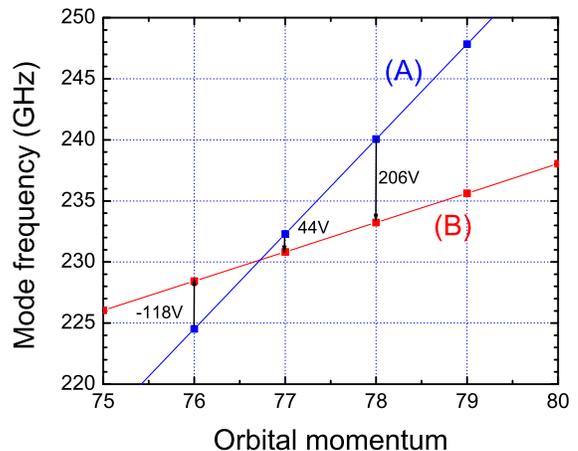}}\vspace*{-0.2in}
\caption[]{\label{fig:phasematchT} Phase-matching solutions for lithium tantalate, Type-II up-conversion, can be achieved at $L_c=76,77,78$ with the DC bias voltages shown on the plot. Curve (A) is the phase-matching curve resulting from Eq.~(\ref{phasematch1}); curve (B) is the microwave dispersion curve calculated numerically. }\vspace*{-0.1in}
\end{figure}

\begin{figure}[b]
\begin{center}
\includegraphics[clip,angle=0,height=5cm]{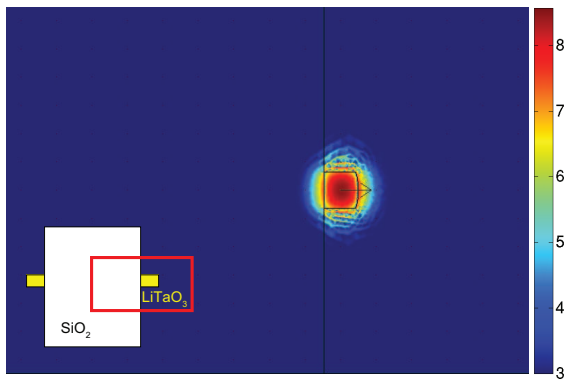}
\caption{Absolute value of the $E_r$-component of a microwave WGM mode in a lithium tantalate resonator with $\omega_c= 2\pi\cdot 238$ GHz, $L_c=80$. The plot is on a log scale with color map given on the right.  A sketch of the geometry is shown in the inset, where the red box represents the calculation window.}
\label{fig:wfRing1}
\end{center}\vspace*{-0.2in}
\end{figure}

The Type-II up-conversion can be realized, however, in weakly birefringent materials. For example, in stoichiometric lithium tantalate the birefringence term is equal to approximately $\pm 2\pi\cdot 0.366$ THz \cite{bruner03}. While the ``positive" solution is still impossible, the ``negative" phase matching solution now can be achieved, see Fig.~\ref{fig:phasematchT}. One important distinction between the Type-I and Type-II configurations is that in the latter the microwave field has to be polarized in the plane of the ring, as required for efficient nonlinear coupling \cite{savchenkov09ssb}. The radial field distribution for one of the microwave WGMs found in Fig.~\ref{fig:phasematchT} is shown in Fig.~\ref{fig:wfRing1}. 

For the numeric simulations shown in Figs.~\ref{fig:phasematchT} and \ref{fig:wfRing1} we used the microwave refraction index of lithium tantalate $n_e\approx n_o=6.5$ \cite{auston88}.
All sizes of this ring were taken the same as for the lithium niobate ring, except the inner radius that was assumed to be 2.61 mm. In these simulations we have not attempted to achieve the integer-valued solution by adjusting the ring sizes.  Instead, we take advantage of the fact that the ordinary and extraordinary optical WGM families can be frequency-tuned relative to each other by temperature, or by bias DC voltage. For voltage tuning, the viable case of Eq.~(\ref{phasematch1}) can be put in the form
\begin{equation}
\omega_c(L_c)\approx\Omega L_c-\omega_0\frac{\Delta n}{n}-\omega_0n^2\frac{r_{33}-r_{31}}{2}\frac{U}{h},\label{phasematch2}
\end{equation}
where the difference of electro-optical constants $r_{33}-r_{31}\approx 22$ pm/V \cite{boyd} determines the phase matching frequency tuning with bias DC voltage $U$ that can be applied to the ring along the axis. Equation (\ref{phasematch2}) allows us to calculate the bias voltages required to achieve the phase matching in three cases shown in  Fig.~\ref{fig:phasematchT}. 

We have already pointed out that the microwave WGMs also can be frequency-tuned by a significant fraction of their FSR. It is interesting to contemplate a possibility of making a broadly-tunable microwave up-converter enabled by the combination of tunable phase-matching (\ref{phasematch2}) and tunable microwave WGMs.  

\section{Conversion efficiency}
To estimate the Type-I conversion efficiency, we use the result by \cite{ilchenko03mod} for the ratio of the optical sidebands powers $P_\pm$ to the pump power $P_0$ in an electro-optical modulator:
\begin{equation}
\frac{P_\pm}{P_0}=\left(\frac{2\xi\sqrt{P_M}}{1+2\xi^2P_M}\right)^2,\quad
\xi=\frac{4gQ}{\omega_0}\sqrt{\frac{Q_M}{\hbar\omega^2_c}},\label{efficiency1}
\end{equation}
where $Q$ and $Q_M$ are the optical and microwave WGM quality factors, respectively, and $P_M$ is the microwave power. We are interested in up-conversion of very low microwave powers, eventually at the single photon levels, which allows us to make the approximation $1+2\xi^2P_M\approx 1$, turn Eq.~(\ref{efficiency1}) around and find the up-conversion efficiency in terms of the photon numbers:
\begin{equation}
\frac{<N_\pm>}{<N_M>}=\frac{P_\pm}{P_M}\frac{\omega_c}{\omega_0}\approx 4\xi^2P_0\frac{\omega_c}{\omega_0}.\label{efficiency2}
\end{equation}

Our numerical simulations allow us to estimate the coupling constant (\ref{g}) and to find the optical pump power $P_0$ required for conversion efficiency (\ref{efficiency2}) to approach unity. Near the unity conversion efficiency the estimate (\ref{efficiency2}) may not be accurate, and saturation effects need to be taken into account. However the purpose of our estimate is to show that nearly absolute conversion efficiency can be achieved even with modest assumptions concerning the system parameters.

For the Type-I up-conversion in lithium niobate, $r_{ij}=r_{33}= 29$ pm/V $= 8.7\cdot 10^{-7}$ esu \cite{boyd}, $n_{a,b} = 2.137$ and $n_c=5.15$. The eigenfunctions of the optical modes are practically equivalent, $\Psi_a\approx \Psi_b$, and the microwave eigenfunction $\Psi_c$ can be treated as a constant and taken out from the integral. The ``mode volume" $V$ in (\ref{g}) is defined as a volume integral of the absolute square of the eigenfunction. Therefore the optical mode volume cancels out, while the microwave mode volume and field amplitude at the optical WGM location are estimated based on the simulation data. To complete the estimate we assume $Q=10^8$ and $Q_M=100$. 
Substituting these numbers into Eq.~(\ref{efficiency2}) we find the photon-number conversion efficiency approaching unity (1/2 for the Stokes, and 1/2 for the anti-Stokes conversion efficiencies) at the optical pump power $P_0\approx$50 mW.

For the Type-II up-conversion in lithium tantalate, $r_{ij}=r_{42}=r_{51}= 20$ pm/V $= 6\cdot 10^{-7}$ esu \cite{boyd}, $n_{a,b} = 2.12$ and $n_c=6.5$, \cite{roberts92,shoji97}. Eq.~(\ref{efficiency1}) in this case remains the same except for the  factor 2 in the denominator, which does not affect the expression for the conversion efficiency (\ref{efficiency2}). The latter approaches unity at $P_0\approx$ 120 mW.

\section{Demonstration of microwave WGMs}

\begin{figure}[b]
\centerline{
\input epsf
\setlength{\epsfxsize}{2.7in} \epsffile{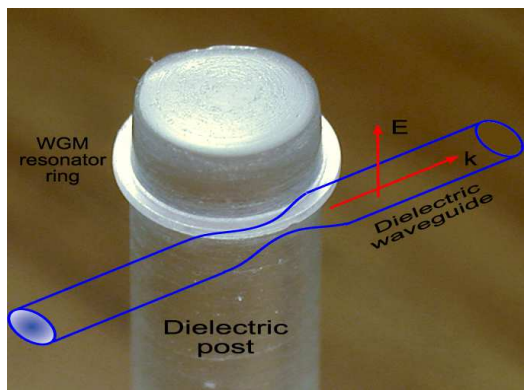}}
\caption[]{\label{fig:schematic2}A lithium niobate WGM ring resonator mounted on a fused silica post (photo) coupled with a tapered dielectric waveguide (drawing). The ring height is 0.29 mm, outer radius is 2.9 mm.}
\end{figure}

As an experimental demonstration, we machined a lithium niobate ring closely matching the parameters calculated above and mounted it on a slightly tapered fused silica post as shown in Fig.~\ref{fig:schematic2}. In this experiment the outer rim of the ring was shaped as a cylinder ($\rho = \infty$) instead of a spheroid, which had little effect on its microwave spectrum. At the mounting location, the outer radius of the silica stem was 2.48 mm. However due to the roughness and tapering caused by the drilling process, the effective inner radius of the lithium niobate ring was slightly larger. In our analysis we treated this radius as a free parameter and found an excellent agreement between the numerical simulations and experimental data for $R_{in}=$ 2.61 mm. 

\begin{figure}[t]\vspace*{-0.1in}
\centerline{
\input epsf
\setlength{\epsfxsize}{3.4in} \epsffile{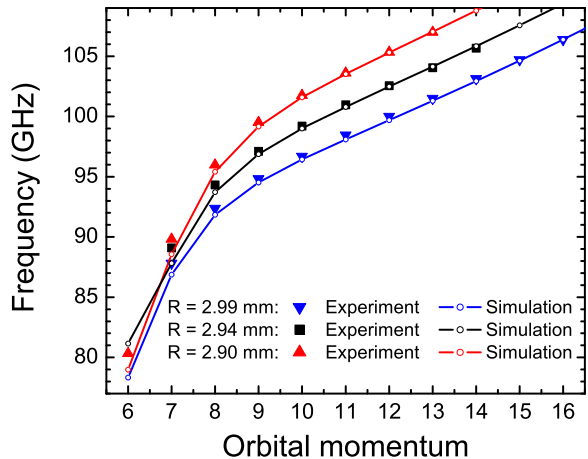}}\vspace*{-0.2in}
\caption{Dependence of the microwave spectrum on the resonators outer radius.  The experimental data and theoretical calculations done by finite element method show very good agreement for the effective inner ring radius determined to be $R_{in}=2.61$ mm.
}\label{fig:ringtuning}
\end{figure}

Microwaves were supplied by a tapered dielectric waveguide coupling to the resonator through the evanescent field. In this experiment we used a fused silica rod stretched over a hydrogen torch to approximately half of its initial diameter of 3 mm. This technique allowed us to achieve the optimal coupling by translating the resonator along the waveguide until their \emph{effective} microwave refraction indices match. 
We gradually polished the disk rim, reducing the outer radius from its initial value of 2.99 mm to 2.90 mm. The microwave dispersion curves were thereby shifted, see Fig.~\ref{fig:ringtuning}. 

\begin{figure}[t]
\centerline{
\input epsf
\setlength{\epsfxsize}{3.3in} \epsffile{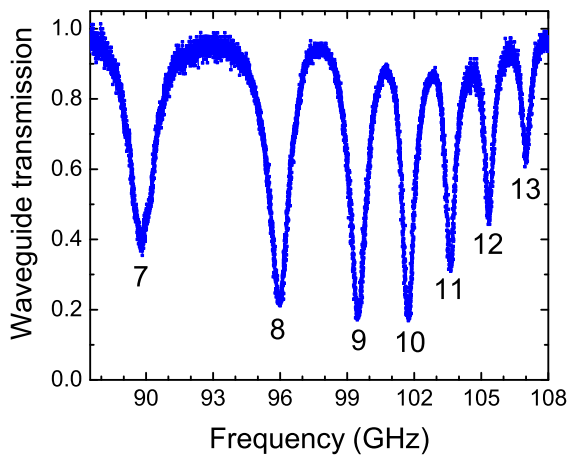}}\vspace*{-0.2in}
\caption[]{\label{fig:ringspectrum} Microwave spectrum of the ring resonator shown in Fig.~\ref{fig:schematic2}. The modes are labeled by the orbital number $L_c$.
}
\end{figure}

The thin ring used in our experiment acts as a looped single-mode microwave waveguide. Therefore the mode sequence in
Fig.~\ref{fig:ringtuning} is indexed by the orbital number $L_c$ determined from the numeric simulation. The simulated dispersion curves are also shown in Fig.~\ref{fig:ringtuning}. 
The microwave
spectrum for the final value $R=2.90$ mm is shown in
Fig.~\ref{fig:ringspectrum}. From this spectrum we see that signal coupling from the THz waveguide into the resonator as high
as 82\% was achieved. We determine the microwave WGM
quality factor to be $Q\approx 100$.

\section{Summary}
To summarize, we have studied microwave WGMs in the ring resonators. The purpose of this study has been to determine the utility of such systems as efficient microwave-to-optics converters, with the focus made on efficient coupling of microwaves into the resonator's WGMs; improving the overlap between the microwave and optical fields; and theoretical demonstration of the phase-matching for desired signal frequency. All of this has been achieved, and we believe that the actual demonstration of nearly unity-efficient microwave-to-optics conversion with subsequent optical photon counting is now feasible.
The benefits from its practical implementation are expected in the
areas of quantum information (e.g., quantum computing with
quantum electronic circuits), astronomy and spectroscopy.

\section{Acknowledgements}
The experimental research described in this paper was carried out at the Jet
Propulsion Laboratory, California Institute of Technology, under a
contract with the NASA.


\begin{thebibliography}{99}

\bibitem{chiou72apl} W. C. Chiou and F. P. Pace,  Appl. Phys. Lett. {\bf 20}, 44 (1972).

\bibitem{abbas76ao} M. M. Abbas, T. Kostiuk, and K. W. Ogilvie,  Appl. Opt. {\bf 15}, 961 (1976).

\bibitem{albota04ol} M. A. Albota and F. N. C. Wong, Opt. Lett. {\bf 29}, 1449 (2004).

\bibitem{karstad05ole} K. Karstad, A. Stefanov, M. Wegmuller, H.
    Zbinden, N. Gisin, T. Aellen, M. Beck, and J. Faist, Opt. Lasers  Engineer. {\bf 43},
    537-544 (2005).

\bibitem{temporao06ol} G. Temporao, S. Tanzilli, H. Zbinden, N.
    Gisin, T. Aellen, M. Giovannini, and J. Faist,  Opt. Lett. {\bf 31}, 1094-1096 (2006).

\bibitem{vandevender07josab} A. P. VanDevender and P. G. Kwiat, J. Opt. Soc. Am. B {\bf 24},
    295-299 (2007).

\bibitem{ding06} Y.J. Ding and Wei Shi, Solid-State Electron. \textbf{50}, 1128 (2006).

\bibitem{khan07} M.J. Khan, J.C. Chen and S. Kaushik, Opt. Lett. \textbf{32}, 3248-50 (2007).

\bibitem{strekalov08THz1} D. V. Strekalov, A. A. Savchenkov, A. B. Matsko and N. Yu, Opt. Lett. 34, 713-715 (2009).

\bibitem{Matsko08THzTheory}  A. B. Matsko, D. V. Strekalov and N. Yu, Phys. Rev. A, \textbf{77}, 043812 (2008).

\bibitem{rabiei02jlt} P. Rabiei, W. H. Steier, C. Zhang, and
    L. R. Dalton,  L.
    Lightwave Technol. {\bf 20}, 1968-1975 (2002).

\bibitem{cohen01el-a} D. A. Cohen and A. F. J. Levi,  Electron. Lett. {\bf 37}, 37-39 (2001).

\bibitem{ilchenko03mod} V. S. Ilchenko, A. A. Savchenkov, A. B.
    Matsko, and  L. Maleki,  JOSA B \textbf{20}, 333-342 (2003).
	
\bibitem{savchenkov09ssb}	A. A. Savchenkov, W. Liang, A. B. Matsko, V. S. Ilchenko, D. Seidel, and L. Maleki, Opt. Lett., \textbf{34}, 1300-1302 (2009).

\bibitem{hosseinzadeh06mtt} M. Hossein-Zadeh, and A. F. J. Levi,
     IEEE Trans. Microwave Theor. and Techniques {\bf 54}, 821-831 (2006).

\bibitem{braginsky87} V.B. Braginsky, V.S. Ilchenko, and
Kh.S. Bagdassarov, Phys. Lett. A, 120, 300-305 (1987).
 
\bibitem{strekalov09lasphys} D. V. Strekalov, A. A. Savchenkov, A. B. Matsko and N. Yu, Laser Phys. Lett., \textbf{6}, 129 (2009).

\bibitem{comsol} COMSOL Multiphysics User's Guide Version 3.5, COMSOL AB, Stockholm, Sweden, 2008.

\bibitem{oxborrow} M.  Oxborrow, IEEE T. Microw. Theory {\bf 55}, 1209-1218 (2007).

\bibitem{palik} ``Handbook of optical constants of solids", ed. E.D.Palik, Acad. Press, New York (1998).

\bibitem{bruner03} A. Bruner, D. Eger, M.B. Oron, P. Blau, M. Katz, and S. Ruschin, Opt. Lett., \textbf{28}, 194-196 (2003).

\bibitem{auston88} D.H. Auston and M.C. Nuss, IEEE J. of Q. El., \textbf{24}, 184-197 (1988).

\bibitem{boyd}``Nonlinear optics", R. Boyd, Acad. Press, New York (2003). 

\bibitem{roberts92} D.A. Roberts, IEEE J. of Q. El., \textbf{28}, 2057-2074 (1992).

\bibitem{shoji97} I. Shoji, T. Kondo, A. Kitamoto, M. Shirane, and R. Ito, JOSA B \textbf{14}, 2268-2294 (1997).

\end{thebibliography}
\end{document}